\documentclass[a4paper,12pt,dvipdfmx]{article}
\usepackage[top=3cm,bottom=3.4cm,left=2cm,right=2cm]{geometry}

\usepackage{bm}
\usepackage{amsmath}
\usepackage{amssymb}
\usepackage{cite}
\usepackage{graphicx}
\DeclareMathAlphabet{\mathpzc}{OT1}{pzc}{m}{it}
\newcommand{\nn}{\nonumber\\}
\newcommand{\bra}[1]{\left<#1 \right|}
\newcommand{\ket}[1]{\left| #1 \right>}

\renewcommand{\thepage}{}
\makeatletter
\@addtoreset{equation}{section}
\renewcommand{\theequation}{\thesection.\@arabic\c@equation}
\makeatother
\renewcommand{\thefootnote}{\fnsymbol{footnote}}

\begin{document}
\begin{titlepage}
\title{
\vspace*{-4ex}
\hfill
\begin{minipage}{3.5cm}
\end{minipage}\\
 \bf 
Reduction of Open String Amplitudes\\
by Mostly BRST Exact Operators
\vspace{0.5em}
}

\author{
Shigenori {\sc Seki}\footnote{\tt sigenori@yukawa.kyoto-u.ac.jp}\medskip\\ 
{\it Osaka City University Advanced Mathematical Institute (OCAMI),}\\
{\it 3-3-138, Sugimoto, Sumiyoshi-ku, Osaka 558-8585, Japan}\bigskip \\
and\bigskip\\
Tomohiko {\sc Takahashi}\footnote{\tt tomo@asuka.phys.nara-wu.ac.jp}\medskip\\
{\it Department of Physics, Nara Women's University,}\\
{\it Nara 630-8506, Japan}
}
\date{}
\maketitle

\begin{abstract}
\normalsize

We study two and three-point tree-level amplitudes of open strings.
These amplitudes are reduced from higher-point correlation functions by
using mostly BRST exact operators for gauge fixing.
For simplicity, we focus on an open string tachyon.  The two-point
amplitude of open string tachyons is reduced from a three-point function
of two tachyon vertex operators and one mostly BRST exact operator.  Similarly the
three-point amplitude is from a four-point function of three tachyon
vertex operators and one mostly BRST exact operator.  One can also obtain the
two-point amplitude from the four-point function of two tachyon vertex operators
and two mostly BRST exact operators.  In these derivation from
four-point functions, moduli integrals are significant.  We discuss the
overall signs of amplitudes which are indefinite in this formalism.

\end{abstract}

\vfill\noindent
\date{12 August 2021}
\end{titlepage}

\renewcommand{\thepage}{\arabic{page}}
\renewcommand{\thefootnote}{\arabic{footnote}}
\setcounter{page}{1}
\setcounter{footnote}{0}

\tableofcontents

\section{Introduction}

String scattering amplitudes is one of important origins of string
theories. It has been studied in detail for a long time and even now
provides unexpected and interesting subjects for us. In bosonic open
string theory, the Veneziano amplitude given by the Euler beta function
was discovered to explain stringy nature such as duality, and in a
modern formulation, more generic string amplitudes are given as
correlation functions of vertex operators in the Polyakov path integral,
where the measure has to be divided by the volume of diffeomorphism and
Weyl symmetry. For open string tree amplitudes, correlarators has to be
divided by the volume of $PSL(2,\mathbb{R})$ on an upper-half plane and
so the one and two-point amplitudes were thought to be vanished due to
the infinite volume of the residual symmetry. However, this view has
been overturned by \cite{Erbin:2019uiz} for two-point amplitudes.  They
have shown that the infinite volume and the divergence of
energy-conserving delta function $\delta(0)$ are canceled each other for
the two-point amplitudes and then the standard free particle expression
can be obtained.

Motivated by their work, the present authors introduced the operator
which fixes a part of the $PSL(2,\mathbb{R})$ gauge symmetry in the BRST
operator formalism.  The operator is defined by
\begin{align}
{\mathpzc E}(z)\equiv \frac{1}{\pi}\int_{-\infty}^\infty dq\,
\Bigl(
c\partial X^0e^{-iqX^0}-i\alpha' q(\partial c)e^{-iqX^0}\Bigr)\,,
\label{eq:B}
\end{align}
where $X^0(z,\bar{z})$ is the 0-th string coordinate and $c(z)$ is the
ghost field, and $z$ is a point on the boundary.\footnote{
In this paper, we multiply the operator ${\cal V}_0$
in \cite{Seki:2019ycz} by the factor $i\alpha'$ to define
the mostly BRST operator ${\mathpzc E}$.} 
This is a BRST invariant operator and moreover it is written by using
the BRST charge $Q_{\rm B}$:
\begin{align}
  {\mathpzc E}(z)= \frac{1}{\pi}\int_{-\infty}^\infty dq\,
\frac{i}{q}\, \delta_B e^{-iqX^0}\,, 
\quad \delta_B e^{-iqX^0} = [\,Q_{\rm B},\,e^{-iqX^0}\,] \,.
\label{eq:V0exact}
\end{align}
We call it a {\it mostly} BRST exact operator, since the integrand for
$q\neq 0$ is indeed BRST exact, but it is not so at the singular point
$q=0$.  In \cite{Seki:2019ycz}, the two-point amplitude is reproduced by
calculating the three point function in which two arbitrary on-shell
open string vertex operators and the operator (\ref{eq:B}) are inserted
on an upper-half plane.

The concept of ``mostly BRST exactness'' is very important.  The
correlation function including the integrand of (\ref{eq:B}) should
behave as a kind of distributions supported at $q=0$, since it becomes
zero for $q\neq 0$ due to BRST exactness. Then, performing the $q$-integration, 
we should have a non-zero result, which is applicable to
two-point amplitudes.  Mostly BRST exactness also assures that the
operator satisfies Lorentz and conformal invariance, although depending
only on the time-direction. Moreover, the concept can be widely applied
in the BRST formalism. Indeed, a mostly BRST exact operator is
constructed in the pure spinor formalism of superstring theory
\cite{Kashyap:2020tgx}, and then two-point superstring amplitudes have
been found to agree with the corresponding expression in the field
theories.

An insertion of the mostly BRST exact operator fixes one degree of
freedom of $PSL(2,\mathbb{R})$, since the operator is a BRST extension
of the gauge fixing delta functional and the Fadeev-Popov determinant
discussed in \cite{Erbin:2019uiz}.  Then, it is natural to ask whether
general $n$-point amplitudes can be obtained by fixing a part of
$PSL(2,\mathbb{R})$ in terms of the mostly BRST exact operator, as
mentioned in the last part of \cite{Seki:2019ycz}. The main purpose of
this paper is to report on the results of the $n$-point amplitudes by
inserting this gauge fixing operator. Consequently, it is confirmed that
the mostly BRST exact operator works well for gauge fixing of
$PSL(2,\mathbb{R})$.

In the derivation of the two-point amplitudes in \cite{Seki:2019ycz},
the energy-conserving delta function becomes $\delta(q)$, if two states
have the same mass and one of them is an incoming state and the other
outgoing. After the $q$-integration, we can find the correct expression
of two-point amplitudes not involving the delta function for energy.
However, if we attempt to evaluate three-point amplitudes by an
additional insertion of the gauge fixing operator, the positions of the
two matter vertex operators is fixed, but the position of the remaining
vertex has to be integrated.  Compared with the conventional three-point
amplitudes, this expression includes, as extra components, the moduli
and $q$ integrations. In this paper, we indicate that the
Feynman $i\varepsilon$ plays a key role in elimination of these
integrals to derive the correct amplitudes.

In the Veneziano amplitude, there are poles for the Mandelstam variables
and, precisely, the poles are defined by the $i\varepsilon$ prescription
as well as Minkowski amplitudes\cite{Polchinski:1998rq}.  This
prescription is related to the fact that the string worldsheet hides
Lorentzian nature although it is treated as Euclidean space
generically\cite{Witten:2013pra}. Similarly, the moduli integral of our
amplitudes generates poles which include the delta function of the
energy zero mode $q$. Then, we will find that, after the
$q$-integration, the conventional expression of three-point amplitudes
is derived from the additional insertion of the gauge fixing vertex.

We should comment on the indefiniteness of the overall sign of the
amplitude given by our gauge fixing operators. First, we remind that 
even for conventional amplitudes in the BRST formalism, the overall sign
is indefinite and it is fixed by referring to the Jacobian factor in the
Fadeev-Popov procedure.\footnote{The sign can also be determined by
deriving the amplitude from string field theories based on the BRST
formalism.} This indefiniteness arises from removing the absolute value
from the Fadeev-Popov determinant to introduce the ghost fields.  The
same extract of the absolute value causes uncertainty in the amplitudes
with our gauge fixing operator. Then, we will see the signature factor
depending on momenta of open string states even in the two-point
amplitude.

More precisely, we will show that conventional open string amplitudes
can be reproduced by using our gauge fixing operator for $PSL(2,\mathbb{R})$,
except the overall sign which can be positive, negative or zero.  The
sign factor is expressed by using momenta of external strings and we may
attempt to interpret this expression in terms of a signed intersection
number associated with the gauge fixing condition. This interpretation
is essentially same as the discussion of a subtlety relating to a gauge
choice in \cite{Eberhardt:2021ynh}, which is inspired by the result of
\cite{Erbin:2019uiz}.

We begin in the section 2 by illustrating the derivation of the two and
three-point amplitudes fixed a part of $PSL(2,\mathbb{R})$ by the
insertion of the mostly BRST exact operator.  For simplicity, we choose
open string tachyons as external states. We then verify in the
subsection 2.3 the overall sign factor by interpreting it as a signed
intersection number.  In the section 3, we fixes two degrees of freedom
in $PSL(2,\mathbb{R})$ by the mostly BRST exact operators to derive
two-point amplitudes for open string tachyons. This result justifies our
gauge fixing procedure of $PSL(2,\mathbb{R})$ and the interpretation of
the overall sign in terms of the intersection.  Finally, we provide
concluding remarks in the section 4.

\section{Reduction by one ${\mathpzc E}$}

\subsection{Two-point amplitude from three-point function}

In \cite{Seki:2019ycz}, it is shown that the mostly BRST exact operator
leads to correct two-point amplitudes for any open string vertex
operators.  Here, we illustrate how to derive a two-point amplitude in
terms of open string tachyons in the 26 dimensional flat Minkowski
spacetime.

We consider an upper-half plane as the world-sheet and the real axis is
regarded as the open string boundary.  If a part of $PSL(2,\mathbb{R})$
is fixed by the insertion of ${\mathpzc E}$, the amplitude for two
tachyon vertex operators is given by
\begin{align}
 {\cal A}_{2}
=ig_{\rm o}^2 C_{D_2} \bra{0}{\mathpzc E}(y_0)\,
cV_1(y_1)\,cV_2(y_2)\ket{0} .
\end{align}
where the tachyon vertex operator $V_i$ is defined as  
\begin{align}
V_i(y) \equiv e^{ip_i\cdot X}(y) \,,
\end{align}
and $C_{D_2}$ is a normalization factor for disk amplitudes:
$C_{D_2}=1/(\alpha' g_{\rm o}^2)$ \cite{Polchinski:1998rq}.  The
positions $y_i$ are along the real axis and we set $y_0<y_1<y_2$.  The
tachyon vertex operators include a factor of the open string coupling
$g_{\rm o}$, but $g_{\rm o}$ is not assigned to the operator ${\mathpzc
E}$, because ${\mathpzc E}$ does not add an extra string to the
process. $\ket{0}$ is the $SL(2,\mathbb{R})$ invariant vacuum normalized
as $\langle 0|0\rangle =(2\pi)^{26}\delta^{26}(0)$.

Calculating the correlation function, we
obtain
\begin{align}
 {\cal A}_{2}&= ig_{\rm o}^2 C_{D_2}{\alpha'(p_1^0-p_2^0) \over i} (2\pi)^{25}
\delta^{25}({\bm p}_1+{\bm p}_2)
\int_{-\infty}^\infty dq
\,\delta(q+p_1^0+p_2^0)
|y_{01}|^{-2\alpha'qp_1^0}
|y_{02}|^{-2\alpha'qp_2^0}|y_{12}|^{\alpha' q^2}\,.
\end{align}
where $y_{ij}$ is defined as $y_{ij}\equiv y_i-y_j$.  The integrand with
the exponent including $q$ implies that the operator in the
$q$-integration of ${\mathpzc E}$ apparently breaks conformal invariance
for $q\neq 0$, however it will be found immediately that the resulting
amplitude has the invariance and no dependence of the position of the
vertex operators.

By the on-shell condition $\alpha' (p_i)^2=1$, we have $p_i^0=\pm
\sqrt{({\bm p}_i)^2+(-1/\alpha')}$ and so the amplitude vanishes if
$p_1^0$ and $p_2^0$ have the same signature due to the factor
$p_1^0-p_2^0$.  In general, this vanishing property is followed from the
BRST invariance, since only the $q=0$ part is not BRST exact in ${\mathpzc E}$ 
and so energies for two tachyons should be conserved in the non-zero
amplitude.  Actually, if $p_1^0$ and $p_2^0$ have opposite sign, we find
that the integral depends only on the contribution from $q=0$, namely
including $\delta(q)$, and then the amplitude becomes non-zero.  In this
case, one tachyon corresponds to an incoming state and another is
outgoing.

The resulting amplitude is given by
\begin{align}
 {\cal A}_{2} =\frac{p_1^0}{|p_1^0|}\times 2|p_1^0|
(2\pi)^{25}\delta^{25}({\bm p}_1+{\bm p}_2)\,.
\label{eq:A32}
\end{align}
Thus, we can obtain the two-point amplitude in the standard free particle
expression, however it should be noted that the signature is not
determined by this gauge fixing procedure, as mentioned in the
introduction. This sign factor is depend on a momentum of external states.

\subsection{Three-point amplitude from four-point function}

Let us consider a three open tachyon amplitude by fixing the two
vertex operators to positions $y_1,\ y_2$ on the real axis.
To fix the residual gauge symmetry of $PSL(2,\mathbb{R})$, we insert the
operator ${\mathpzc E}$ to a position $y_0$.
\begin{align}
 {\cal A}_{3} =
i g_{\rm o}^3 C_{D_2}\bra{0}{\mathpzc E}(y_0)\,
cV_1(y_1)\,cV_2(y_2)\int_{-\infty}^\infty dy_3\,V_3(y_3)\ket{0}. \label{eq:3ptamp}
\end{align}
Although it is a three point amplitude, the position $y_3$ of the third
vertex operator is integrated as a result of the insertion of ${\mathpzc E}$. 
Since $y_3$ varies from $-\infty$ to $\infty$, this expression
includes two ordering of the three tachyons and so it is not needed to
add an extra contribution from the exchange, $p_1\leftrightarrow p_2$,
unlike a conventional case in \cite{Polchinski:1998rq}.

Here, for convenience, introducing a covariant expression for the
momentum $q$ in $\mathpzc E$ defined by \eqref{eq:V0exact}:
$p_0=(q,0,\dots,0)$, we rewrite $\delta_{\rm B} e^{-iqX^0}$ as
$\delta_{\rm B}e^{ip_0\cdot X}$.  We fix the positions as $y_0<y_1<y_2$,
since the ordering of $y_0$, $y_1$ and $y_2$ is related only to the
overall sign.  The correlation function in the amplitude
\eqref{eq:3ptamp} can be calculated as
\begin{align}
{\cal F}_1(q)&= \bra{0}\biggl({i \over \pi q}\delta_{\rm B}e^{ip_0\cdot X}(y_0)\biggr)\,cV_1(y_1)\,cV_2(y_2)
\int_{-\infty}^\infty dy_3\,V_3(y_3)\ket{0} \nn
&= \biggl|\frac{y_{01}y_{02}}{y_{12}}\biggr|^{-\alpha'(p_0)^2}
(2\pi)^{26}\delta^{26}(\sum_{j=0}^3 p_j)\, I \,, 
\nn
I &= {i \over \pi q}\int_{-\infty}^\infty dx\,
\frac{1}{2}\biggl(\alpha'(p_0)^2+2\alpha'p_0\cdot p_1-2\alpha'p_0\cdot p_3\frac{x}{1-x}\biggr) 
|x|^{2\alpha'p_2\cdot p_3}|1-x|^{2\alpha'p_0\cdot p_3} \,, \label{eq;modint}
\end{align}
where $x$ denotes a moduli parameter given by the cross ratio,
\begin{align}
 x=\frac{y_{01}y_{23}}{y_{02}y_{13}}\,.
\end{align}
The prefactor with the exponent $-\alpha'(p_0)^2=\alpha' q^2$ implies
that ${\mathpzc E}$ apparently breaks $PSL(2,\mathbb{R})$ invariance for $q\neq 0$,
but it is not problematic since the final result should depend only on
the $q=0$ part.

The amplitude is written in terms of the Mandelstam variables;
\begin{align}
 s=-(p_0+p_1)^2,\quad t=-(p_0+p_2)^2,\quad u=-(p_0+p_3)^2 .
\end{align}
However, unlike the Veneziano amplitude, these are independent since
$p_0$ does not satisfy an on-shell condition, {\it i.e.}, $s+t+u=q^2-3/\alpha'$. 
The moduli integral \eqref{eq;modint} is rewritten as 
\begin{align}
I(q) = {-i \over 2\pi q}\int_{-\infty}^\infty dx \biggl( (\alpha's+1){1 \over 1-x} +(\alpha't+1){x \over 1-x} \biggr) 
	|x|^{-\alpha's-2} |1-x|^{\alpha'(s+t)+2}\,.
\end{align}
Noting that the moduli integral $I$ splits into three ranges, {\it i.e.}, $I = I_{\rm (i)} + I_{\rm (ii)} +I_{\rm (iii)}$, for 
\begin{align}
 ({\rm i})\  -\infty<y_3<y_0,\ y_2<y_3<\infty\qquad
 ({\rm ii})\ y_0<y_3<y_1\qquad
 ({\rm iii})\  y_1<y_3<y_2\,,
\label{eq:regions}
\end{align}
or equivalently  (i) $0<x<1$, (ii) $1<x<\infty$, (iii) $-\infty < x < 0$, 
the amplitude becomes
\begin{align}
{\cal A}_3 = ig_{\rm o}^3 C_{D_2} \int_{-\infty}^\infty dq\, {\cal F}_1
= i g_{\rm o}^3 C_{D_2} \int_{-\infty}^\infty dq\, 
	\biggl|{y_{01}y_{02} \over y_{12}}\biggr|^{-\alpha'(p_0)^2} 
	(2\pi)^{26}\delta^{26}(\sum_{j=0}^3 p_j) 
(I_{\rm (i)} + I_{\rm (ii)} + I_{\rm (iii)}) \,, \label{eq:3ptampBeta}
\end{align}
where 
\begin{align}
I_{\rm (i)} &= {-i \over 2\pi q}\{(\alpha's+1)B(-\alpha's-1, \alpha'(s+t)+2) +(\alpha' t+1)B(-\alpha's, \alpha'(s+t)+2)\} \,, \\
I_{\rm (ii)} &= {-i \over 2\pi q}\{-(\alpha's+1)B(-\alpha't, \alpha'(s+t)+2) -(\alpha't+1)B(-\alpha't-1, \alpha'(s+t)+2)\} \,, \\
I_{\rm (iii)} &= {-i \over 2\pi q}\{(\alpha's+1)B(-\alpha's-1, -\alpha't) -(\alpha't+1)B(-\alpha's, -\alpha't-1)\} \,.
\end{align}
The Euler beta function is defined by
\begin{align}
 B(u,v)=\int_0^1 dx\,x^{u-1}(1-x)^{v-1} \,.
\end{align}

If the momentum variables are in the convergence region of the integral
$B$, we find that $I_{\rm (i)}$, $I_{\rm (ii)}$ and $I_{\rm (iii)}$ vanish. 
This reflects the mostly
BRST exactness of ${\mathpzc E}$, however, it does not necessarily imply
that the amplitude becomes trivially zero. To derive the physical
result, we should study the $q=0$ part of the amplitude, which is given
as the divergence of the integral.

To study the divergence, we have to introduce a convergence factor which
corresponds to the Feynman $i\varepsilon$. For instance, one of the
integrals in $I_{\rm (i)}$ has the convergence factor
$\exp(-i\varepsilon(\log x+\log (1-x)))$ \cite{Witten:2013pra} and it
should be precisely defined as
\begin{align}
 B(-\alpha's-1,\alpha'(s+t)+2)=\int_0^1 dx x^{-\alpha's-2-i\varepsilon}
(1-x)^{\alpha' (s+t)+1-i\varepsilon} \,.
\end{align}
To extract the singularity at $q=0$, we have only to expand it
in rational fractions and to use the formula,
\begin{align}
 \frac{1}{x-i\varepsilon}={\rm P}\frac{1}{x}+\pi i\delta(x) \,,
\end{align}
where P denotes the principal value. In this example, the singularity can
be evaluated as
\begin{align}
B(-\alpha's-1,\alpha'(s+t)+2)
&= \frac{1}{-\alpha' s-1-i\varepsilon} +\frac{1}{\alpha'(s+t)+2-i\varepsilon}+\cdots
\nn
&=
\frac{\pi i}{2\alpha'} \biggl(\frac{1}{|p_1^0|} +\frac{1}{|p_3^0|}\biggr)\delta(q)+\cdots \,.
\label{eq:Fexample}
\end{align}
These terms are generated from the singular configuration of the worldsheet
with $y_3$ approaching to $y_0$ or $y_2$. 

Similarly, by extracting $\delta(q)$, the moduli integration gives the
following results for each integration range:
\begin{align}
I_{\rm (i)} \sim {1 \over 2}\biggl(
\frac{p_1^0}{|p_1^0|} -\frac{p_3^0}{|p_3^0|} \biggr)\delta(q) \,, \quad 
I_{\rm (ii)} \sim {1 \over 2}\biggl(
-\frac{p_2^0}{|p_2^0|} +\frac{p_3^0}{|p_3^0|} \biggr)\delta(q) \,,\quad 
I_{\rm (iii)} \sim {1 \over 2}\biggl(
\frac{p_1^0}{|p_1^0|} -\frac{p_2^0}{|p_2^0|} \biggr)\delta(q) \,,
\end{align}
where $\sim$ stands for equal up to analytic terms. 
By substituting these results into \eqref{eq:3ptampBeta} and performing the $q$-integration,
we obtain the final expression of the amplitude:
\begin{align}
 {\cal A}_{3}=
{1 \over 2}\biggl(\frac{p_1^0}{|p_1^0|}
-\frac{p_2^0}{|p_2^0|}\biggr)\times
\frac{2ig_{\rm o}}{\alpha'}(2\pi)^{26}\delta^{26}(p_1+p_2+p_3) \,.
\end{align}
This result agrees with the correct three string amplitude 
except the prefactor taking the value $\pm 1$ or $0$,
which can not be determined in this gauge fixing as mentioned above.

\subsection{Overall sign of amplitudes}

The resulting three-point amplitude includes the sign factor
\begin{align}
 \frac{1}{2}
\biggl(\frac{p_1^0}{|p_1^0|} -\frac{p_2^0}{|p_2^0|}\biggr) \,,
\label{eq:sign}
\end{align}
which is $\pm 1$ if $p_1^0$ and $p_2^0$ have opposite signs each other,
and becomes zero if they are the same signs.  Let us verify the validity
of this factor.

The operator ${\mathpzc E}$ corresponds to the gauge fixing
$X^0(y_0)=0$, where the point $y_0$ is on the segment of the boundary of
the disk.  In the case $y_0<y_1<y_2$, the segment consists of the
regions, $-\infty < y < y_1$ and $y_2 < y < \infty$, noting that $\pm
\infty$ are identified.  The tachyon vertex operators are inserted at
the points $y_1$ and $y_2$.

From the path integral formulation, we integrate all over the function
$X^0(y)$ on the boundary to derive the amplitude. However, $X^0(y)$
should approach $+\infty$ at the inserted points of the vertex operators
if the open string corresponds to outgoing states, and $-\infty$ for
incomming states, because $X^0$ is the time coordinate of the target space.

Suppose that the operator at $y_2$ is incoming and the one at $y_1$ is
outgoing, which implies $p_2^0>0$ and $p_1^0<0$.  In this case,
$X^0(y_2)=-\infty$ and $X^0(y_1)=+\infty$. Since the equation $X^0(y)=0$
has necessarily roots, it is possible to impose the gauge fixing
condition $X^0(y_0)=0$.
\begin{figure}[h]
\begin{center}
\includegraphics{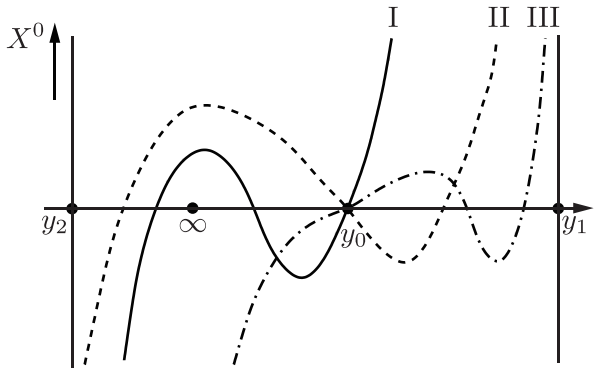} 
\end{center}
\caption{The function $X^0(y)$ with $y$ in the region, 
$-\infty < y < y_1$ or $y_2 < y < \infty$,
for the case that the total number of roots is three.}
\label{fig:X0}
\end{figure}
In Fig.~\ref{fig:X0}, we show the function $X^0(y)$ satisfying the gauge
choice $X^0(y_0)=0$ for the case that the total number of roots is
three.  For the three roots, there are three possibilities how to choose
intersections at $y_0$, namely, I, II and III in Fig.~\ref{fig:X0}.  As
mentioned in the introduction, the operator ${\mathpzc E}$ originates in
the Fadeev-Popov determinant for the gauge fixing, but we have to define
${\mathpzc E}$ by using the singed determinant. Therefore, for the three
possibilities, $\partial X^0(y_0)$ has different signatures: positive
for I and III and negative for II. Then, the amplitude becomes positive
for I and III and negative for II.  Consequently, after summing over all
$X^0$ in the case of the three intersection, the amplitude is given as a
positive result if I, II and III lead to the same absolute value.  So,
we expect that the sign of the amplitude is given as the signed
intersection number of the graph $u=X^0(y)$ with $u=0$ for any number of
the roots.

The signed intersection number is $+1$ for $p_2^0>0$ and $p_1^0<0$, and
it is $-1$ for $p_2^0<0$ and $p_1^0>0$.  If $p_2^0$ and $p_1^0$ have the
same sign, both of the tachyon states are incomming or outgoing and so
the signed intersection number becomes zero. These results agree with
the signature of (\ref{eq:sign}).

There is the case where the point $y_3$ enters the same segment of $y_0$,
but the signed intersection number is unchanged regardless of whether
$X^0(y_3)$ is infinity or minus infinity.

The two point amplitude is given by (\ref{eq:A32}), which is given by
the gauge choice $X^0(y_0)=0$ and the vertex insertions at $y_1$ and
$y_2$. So, we apply the same discussion as the above to the signature of (\ref{eq:A32}). The only difference is that the sum of $p_1^0$
and $p_2^0$ is zero for the two point amplitude, and so the signature
(\ref{eq:sign}) equals to $p_1^0/|p_1^0|$, which agrees with the sign
of (\ref{eq:A32}).

Thus, we can verify the validity of the signature of the amplitude
from the point of view of the signed intersection number.

\section{Reduction by two ${\mathpzc E}$'s}

\subsection{Two-point amplitude from four-point function}

Let us consider two open tachyon amplitudes by inserting the mostly BRST
exact operators at two points, $y_0$ and $y_1$.  To fix $PSL(2,\mathbb{R})$ gauge
symmetry, we fix the one tachyon vertex operator at the position $y_2$
and so the position $y_3$ of another vertex operator is integrated.
The amplitude is given by
\begin{align}
 {\cal A}_{2}&=
\frac{1}{2}\times ig_{\rm o}^2 C_{D_2} \bra{0}
{\mathpzc E}(y_0){\mathpzc E}(y_1) cV_2(y_2)
\int_{-\infty}^\infty dy_3 V_3(y_3)\ket{0}.
\label{eq:A42}
\end{align}
Here, the two mostly BRST exact operators at $y_0$ and $y_1$ are
indistinguishable each other and so we should divide the amplitude by
the statistical factor $2$ from the perspective of the Feynman diagram
of four-point amplitudes.  This factor can be also understood in
world-sheet language. The metric on the upper-half plane is given by
\begin{align}
 ds^2=\frac{dz d\bar{z}}{|z-\bar{z}|^2}\,,
\end{align}
which is invariant under $PSL(2,\mathbb{R})$ transformation:
$z\rightarrow (az+b)/(cz+d)$, $ad-bc=1$. In addition, the metric is
invariant under the discrete transformation $z\rightarrow -\bar{z}$.
These symmetries generate the conformal Killing group (CKG).  In the
usual gauge fixing, the positions of three different vertex operators
are fixed and so any conformal Killing invariance is not remained.
However, if the two positions are fixed by the indistinguishable
operators, a $\mathbb{Z}_2$ invariance as a part of the CKG is
unfixed.
More explicitly, the $\mathbb{Z}_2$
transformation is given by
\begin{align}
 z \rightarrow f(z)=\frac{-(y_0y_1-y_2^2)\bar{z}+y_2(2y_0y_1-y_0y_2
-y_1y_2)}{
-(y_0+y_1-2y_2)\bar{z}+(y_0y_1-y_2^2)}\,.
\end{align}
This mapping function satisfies $f(y_0)=y_1$, $f(y_1)=y_0$, $f(y_2)=y_2$
 and $f(f(z))=z$.  Therefore, the $\mathbb{Z}_2$ symmetry should be
 fixed by restricting the integration range of $y_3$ to half of the
 boundary, or by dividing the factor 2 while the $\mathbb{Z}_2$ symmetry
 remains unfixed.  In the expression (\ref{eq:A42}), we adopt the latter
 procedure to fix the CKG.\footnote{For the Veneziano amplitude, this
 $\mathbb{Z}_2$ symmetry is remained if two momenta are identical,
 however the identical case has measure zero and so this factor is
 irrelavent to the amplitude at generic momenta.}

Now, let us evaluate the correlation function in (\ref{eq:A42})
according to the discussion in the previous section. After tedious
calculation, we obtain
\begin{align}
{\cal F}_2(q,q')& = \bra{0}
\biggl({i \over \pi q}\delta_{\rm B} e^{ip_0\cdot X}(y_0)\biggr)
\biggl({i \over \pi q'}\delta_{\rm B} e^{ip_1\cdot X}(y_1)\biggr)
ce^{ip_2\cdot X}(y_2)
\int_{-\infty}^\infty dy_3 e^{ip_3\cdot X}(y_3) \ket{0} \nn
&= \Big|\frac{y_{12}}{y_{01}y_{02}}\Big|^{\alpha'(p_0)^2}
\Big|\frac{y_{02}}{y_{01}y_{12}}\Big|^{\alpha'(p_1)^2}
(2\pi)^{26}\delta^{26}(\sum_{j=0}^3 p_j) \,J \,, 
\nn
J &= {1 \over \pi^2 qq'}\int_{-\infty}^\infty dx\,\frac{1}{2}\biggl(
\alpha'p_0\cdot p_1 
-(\alpha')^2 (p_0+p_1)^2(p_0\cdot p_1)
\nn
&\qquad
-(\alpha')^2\bigl((p_0)^2+2p_0\cdot p_1\bigr)(p_1\cdot p_3)x
+(\alpha')^2\bigl((p_1)^2+2p_0\cdot p_1\bigr)(p_0\cdot p_3)\frac{x}{1-x}
\nn
&\qquad
+2(\alpha')^2 (p_0\cdot p_3) (p_1\cdot p_3)\frac{x^2}{1-x}
\biggr)
|x|^{2\alpha' p_2\cdot p_3}|1-x|^{2\alpha' p_0\cdot p_3} \,,
\label{eq:A42cf}
\end{align}
where $p_0=(q,0,\cdots,0)$, $p_1=(q',0,\cdots,0)$.  $q$ and $q'$ are
integrated later in ${\mathpzc E}(y_0)$ and ${\mathpzc E}(y_1)$,
respectively.  Due to the BRST invariance, the $q\neq 0$ part does not
contribute to the amplitude and so it can be easily seen that the
$x$-integration of (\ref{eq:A42cf}) vanishes in the convergence region.
Then we have only to extract the singularity of $q=0$ as the previous
case.

We should comment on a different type of singularity proportional to
$\delta((q+q')^2)$, which seems to arise from (\ref{eq:A42cf}), since
there are some terms in the moduli integration which have the pole at $2
\alpha'p_2\cdot p_3+2=-\alpha'(q+q')^2=0$. This pole corresponds to a
massless open string state as the intermediate one of the two tachyon
collision. However, this pole should vanish since the amplitude has
twist symmetry for two tachyons. Indeed, gathering together factors of
this pole, using on-shell condition and momentum conservation, we find
that they totally cancel each other and then the dangerous distribution
does not appear in the amplitude.

Now, we show the final expression of the amplitude by extracting 
$\delta(q)$ and $\delta(q')$. Taking $y_0<y_1<y_2$,
the integration (\ref{eq:A42cf}) for three regions of (\ref{eq:regions}) 
leads to $J = J_{\rm (i)} +J_{\rm (ii)} +J_{\rm (iii)}$ with 
\begin{align}
J_{\rm (i)} \sim \frac{i\alpha'}{2\pi}|p_3^0|\,\delta(q) \,,\quad
J_{\rm (ii)} \sim \frac{i\alpha'}{2\pi}|p_3^0|\,\bigl(\delta(q)+\delta(q')\bigr) \,,\quad 
J_{\rm (iii)} \sim \frac{i\alpha'}{2\pi}|p_3^0|\,\delta(q') \,.
\end{align}
Unless $p_2^0+p_3^0=0$, the amplitude vanishes as well as the two-point
amplitude reduced from the three-point function. Finally, in the case of
$p_2^0+p_3^0=0$, the resulting amplitude is given by
\begin{align}
{\cal A}_{2} 
= {ig_{\rm o}^2 C_{D_2} \over 2}\int_{-\infty}^\infty dq \int_{-\infty}^\infty dq'\, {\cal F}_2 
= -1\times 2|p_3^0|(2\pi)^{25}
\delta^{25}({\bm p}_2+{\bm p}_3) \,. 
\label{eq:A42final}
\end{align}

Similar to the previous cases, the signature of the amplitude can not be
determined in this gauge fixing. However, contrastingly, the signature
of (\ref{eq:A42final}) does not depend on momenta of the vertex
operators.  This can be also understood in terms of the signed
intersection.  For simplicity, suppose that $p_2^0>0$ and $p_3^0<0$, and
so $X^0(y_2)=-\infty$ and $X^0(y_3)=+\infty$. We set $y_0<y_1<y_2$.  If
$y_1<y_3$ or $y_3<y_0$, the intersection number of $X^0$ on the
segment in which $y_0$ and $y_1$ are located becomes an odd number.  By
assigning these intersections to $y_0$ and $y_1$ and adding the
contributions of all combination, the corresponding amplitude is
expected to be zero since $\sum {\rm sgn}(\partial X^0(y_0))
{\rm sgn}(\partial X^0(y_1))=0$ in this case. If $y_0<y_3<y_1$, the
intersection number on each segment for $y_0$ and $y_1$ is odd and
moreover we find that $\sum {\rm sgn}(\partial X^0(y_0))=+1$ and $\sum
{\rm sgn}(\partial X^0(y_1))=-1$. Then, the signature factor is
$+1\times (-1)=-1$ in this case.  It can be easily seen that the same
signature is obtained for $p_2^0<0$ and $p_3^0>0$, and therefore the
signature is independent of momenta in the two-point amplitude reduced
from the four-point function.

\section{Concluding remarks}

We have shown that the two and three-point tree-level amplitudes of open
string tachyons are reduced from the three and four-point correlation
functions by inserting the mostly BRST exact operators ${\mathpzc E}$.
Firstly we have obtained the two-point amplitude by calculating the
three-point correlation function with one ${\mathpzc E}$: $\langle
{\mathpzc E}\, cV_1\, cV_2\rangle$.  This two-point amplitude has the
same expression as the standard free particle, but its overall sign
depends on the sign of the momentum of external state.

Similarly we have shown that the three-point amplitude is reduced from
the four-point function of three tachyon vertex operators and one ${\mathpzc
E}$: $\langle{\mathpzc E}(y_0)\,cV_1(y_1)\,cV_2(y_2)\,\int dy_3
V_3(y_3)\rangle$.  In this computation, the moduli integration plays an
important role, so that $\delta(q)$ appears from the singularities of
the Euler beta functions.  Combining it with the $q$-integration, the
four-point function leads to the correct three-point amplitude up to the
overall sign.  The operator ${\mathpzc E}$ implies the gauge fixing,
$X^0(y_0)=0$.  $X^0(y)$ at $y_1$ and $y_2$, where tachyon vertex operators are
inserted, are equal to $\pm \infty$, of which sign depends on the choice
of incoming or outgoing for each external state.  The configuration of
$X^0(y)$ with $X^0(y_0)=0$ and $X^0(y_{1,2}) = \pm \infty$ fixes the
total signed intersection number of the graph $u=X^0(y)$ with $u=0$.
One can recognize that the overall sign of the two-point amplitude is
given by this intersection number.

We have also studied the two insertions of ${\mathpzc E}$. 
The four-point function $\langle{\mathpzc E}\,{\mathpzc E}\,cV_1\, \int V_2\rangle$ is 
reduced to the two-point amplitude. 
The two ${\mathpzc E}$'s provide $q$ and $q'$-integrals. 
In the similar way as the computation for the reduction from four-point to three, 
the moduli integration yields $\delta(q)$ and $\delta(q')$. 
Due to the $q$ and $q'$-integration with these delta functions, 
we can obtain the correct two-point amplitude. 
There again appear a problem of the overall sign, but in this case 
the signature does not depend on the external momenta. 

In this paper, by focusing only on open string tachyons, 
we have explicitly evaluated the correlation functions, 
and, as a result, we have found they are reduced to the lower-point amplitudes. 
In the case of two-point amplitudes of general vertex operators, 
Refs.~\cite{Seki:2019ycz,Erbin:2019uiz} have shown that 
the two-point amplitudes can be derived from the three-point function. 
However, for the case of higher-point amplitudes, we should understand more 
how the operator ${\mathpzc E}$ acts in the higher-point functions of general vertex operators. 

It is not straightforward to apply our formalism using the operator ${\mathpzc E}$ 
for closed string amplitudes. 
For instance, if we extend ${\mathpzc E}$ to ${\mathpzc E}_{\rm closed} \sim \int {dq \over q} [\,Q^{\rm c}_{\rm B},\, e^{-iqX^0(z,{\bar z})}\,]$, 
where $Q^{\rm c}_{\rm B}$ is the BRST charge of closed string, 
it causes a problem of ghost number. 
Therefore, we need to find a new operator, which plays a same role as ${\mathpzc E}$ 
in the closed string theory \cite{ST}.

\section*{Acknowledgments}
The authors would like to thank Isao Kishimoto for valuable comments.
This work was supported in part by JSPS
Grant-in-Aid for Scientific Research (C) \#20K03972.
S.~S.~was supported in part by MEXT Joint
Usage/Research Center on Mathematics and Theoretical Physics at OCAMI 
and by JSPS Grant-in-Aid for Scientific Research (C) \#17K05421.


\end{document}